\documentclass[preprint,12pt]{elsarticle}

\usepackage{amssymb}

\journal{Physica C}

\begin{document}

\begin{frontmatter}



\title{Effect of Microwave Irradiation on Parametric Resonance in
Intrinsic Josephson Junctions}


\author{Mahmoud Gaafar~$^{a,b}$}
\author{Yury Shukrinov~$^{a,c}$}
\address{$^{a}$ BLTP, JINR, Dubna, Moscow Region, 141980, Russia}
\address{$^{b}$Department of Physics, Faculty of Science, Menoufiya University, Egypt}
\address{$^{c}$ Max-Planck-Institute for the Physics of Complex Systems, 01187 Dresden, Germany}
\begin{abstract}
The effect of microwave irradiation on the phase dynamics of intrinsic Josephson junctions in high temperature superconductors is investigated. We demonstrate the influence of microwave's amplitude variation on the current-voltage characteristics and on the time dependence (temporal oscillations) of the electric charge in the superconducting layers. A remarkable changing of the longitudinal plasma wavelength at parametric resonance is shown. We demonstrate an effect of the microwave radiation on the width of the parametric resonance region.
\end{abstract}

\end{frontmatter}


\section{Introduction}
Intrinsic Josephson junctions (IJJs) in high temperature superconductors (HTSC) under microwave (MW) irradiation have attracted growing interest because of their rich physics and possible applications. IJJs are very complex, and their electrical and magnetic properties are strongly nonlinear. Hence a complete understanding of their physical dynamics is needed today. One of the most spectacular indications of the Josephson effect is locking of Josephson oscillations to the frequency of external MWs. As the Josephson frequency is proportional to voltage, the locking leads to appearance of steps at quantized voltages  in the current-voltage characteristic (CVC), called Shapiro steps (SS) \cite{Shapiro,Tinkham}. These steps appear in the CVC at discrete voltage values $V_n=nh\omega/2e$, where $\omega$ is the frequency of the applied MW signal and $n$ is integer number.

It was shown by Koyama and Tachiki \cite{koyama96} that the system of equations for capacitively coupled
Josephson junctions has a solution corresponding to the longitudinal plasma wave (LPW) propagating along the c-axis. So, the Josephson oscillations can excite the LPW by their periodical actions. The frequency of Josephson oscillations $\omega_J$ is determined by the voltage value in the junction, and at
$\omega_J=2 \omega_{LPW}$, where $\omega_{LPW}$ is LPW frequency, the parametric resonance (PR) is realized. 

In this article we investigate the influence of external radiation on the parametric resonance of IJJs by: i) decrease of frequency of radiation which brings SS near PR region (PRR); ii) increase of amplitude of radiation. Bringing SS near PRR causes the width of this region to be shrink. An increase in the amplitude of radiation leads to (a) changing the wavelength of LPWs along the stack, (b) appearance of additional PR above SS. So, the external radiation essentially changes the physical picture of the coupled JJs.
\section{Model and Methods}
To investigate the phase dynamics of IJJ we use the one-dimensional CCJJ+DC model with the gauge-invariant phase differences $\varphi_l(t)$  between S-layers $l$ and $l+1$ in the presence of electromagnetic irradiation described by the system of equations:
\begin{equation}
\label{syseq} \left\{\begin{array}{ll} \displaystyle\frac{\partial \varphi_{l}}{\partial
t}=V_{l}-\alpha(V_{l+1}+V_{l-1}-2V_{l})
\vspace{0.2 cm}\\
\displaystyle \frac{\partial V_{l}}{\partial t}=I-\sin \varphi_{l}-\beta\frac{\partial \varphi_{l}}{\partial t} + A\sin\omega t + I_{noise}
\end{array}\right.
\end{equation}
where $t$ is dimensionless time, normalized to the inverse plasma frequency $\omega^{-1}_p$, $\omega_{p}=\sqrt{2eI_c/\hbar C}$, $\beta=1/\sqrt{\beta_{c}}$, $\beta_{c}$ is McCumber parameter, $\alpha$  gives the coupling between junctions \cite{koyama96},
$A$ is the amplitude of the radiation.  To find the IVC of the stack of IJJ we solve the system of nonlinear second-order differential equations (1) using the fourth order Runge-Kutta method. In our simulations we measure the voltage in units of $V_0=\hbar\omega_p/(2e)$, the frequency in units of $\omega_{p}$, the bias current $I$ and the amplitude of radiation $A$ in units of $I_c$.

The time dependence of the electric charge in the S-layers is studied by Maxwell equation $\emph{div} (\varepsilon\varepsilon_0 E) = Q$, where $\varepsilon$ and
$\varepsilon_0$ are relative dielectric and electric constants, respectively. The charge density $Q_l$ in the S-layer $l$ is proportional to the difference between the voltages $V_{l}$ and $V_{l+1}$ in the neighbor Josephson junctions  $Q_l=Q_0 \alpha (V_{l+1}-V_{l})$,
where $Q_0 = \varepsilon \varepsilon _0 V_0/r_D^2$.  To see clearly the effect of radiation only, we have taken even number of junctions in the stack $N=10$ to escape the additional modulations of the electric charge oscillations in the PRR which appear in case of odd number of Josephson junctions. Numerical calculations have been done for a stack with the coupling parameter $\alpha = 1$, dissipation parameter $\beta= 0.2$ and periodic boundary conditions. We note that the qualitative results are  not sensitive to these  parameter values and boundary conditions. In our simulations we put $T_m=1000$ for time domain, $\delta t=0.05$ for step in time, $\delta I=0.0001$ for step in current and we add to the bias current a small noise with amplitude in the interval $(+10^{-8}, -10^{-8})$. The details concerning the numerical procedure are given in Ref. \cite{smp-prb07}.
\section{Results and Discussions}
It is known that in case of single JJ an increase in radiation amplitude $A$ decreases a
hysteresis region, i.e., it leads to the decrease of the critical current value and the increase of the return
current $I_R$ \cite{kleiner-book}. For the stack of coupled JJ the external radiation leads additionally to the series of novel effects related to the parametric resonance and the LPW  propagating along the c axis \cite{koyama96,kleiner2000}. We will demonstrate below the changing of LPW wavelength and additional resonances around SS.
\begin{figure}[!ht]
 \centering
\includegraphics[height=70mm]{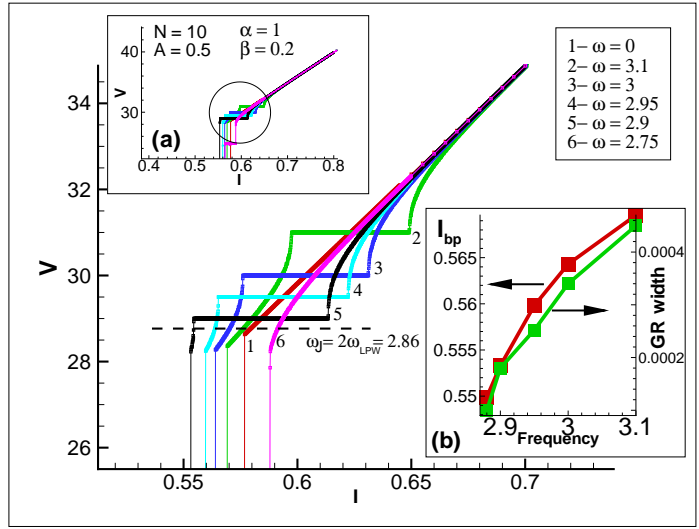}
\caption{CVC for the stack with 10 JJs under radiation with amplitude $A=0.5$ and different frequencies. Curve 1 show CVC without irradiation. Inset (a) stresses the coincidence of the curves before Shapiro step. Inset (b) demonstrates the changing of both breakpoint current ($I_{bp}$) and the growing region (GR) width with frequency variation.}
 \label{1}
\end{figure}
In Fig.~\ref{1} we show the CVC for a stack with 10 JJs at $\alpha=1$ and $\beta=0.2$ under radiation with amplitude $A=0.5$ and different frequencies. Curve 1 show CVC without irradiation. Inset (a) stresses the coincidence of the curves before SS. The SS does not appear at
frequency smaller than $\omega= 2.86$, because before it a jump to another branch is happened. Inset (b) illustrates that, with decreasing the frequency of external radiation and approaching the parametric resonance condition, both the breakpoint current ($I_{bp}$) and the growing region (parametric resonance region) width are decreasing.
\begin{figure}[!ht]
 \centering
\includegraphics[height=70mm]{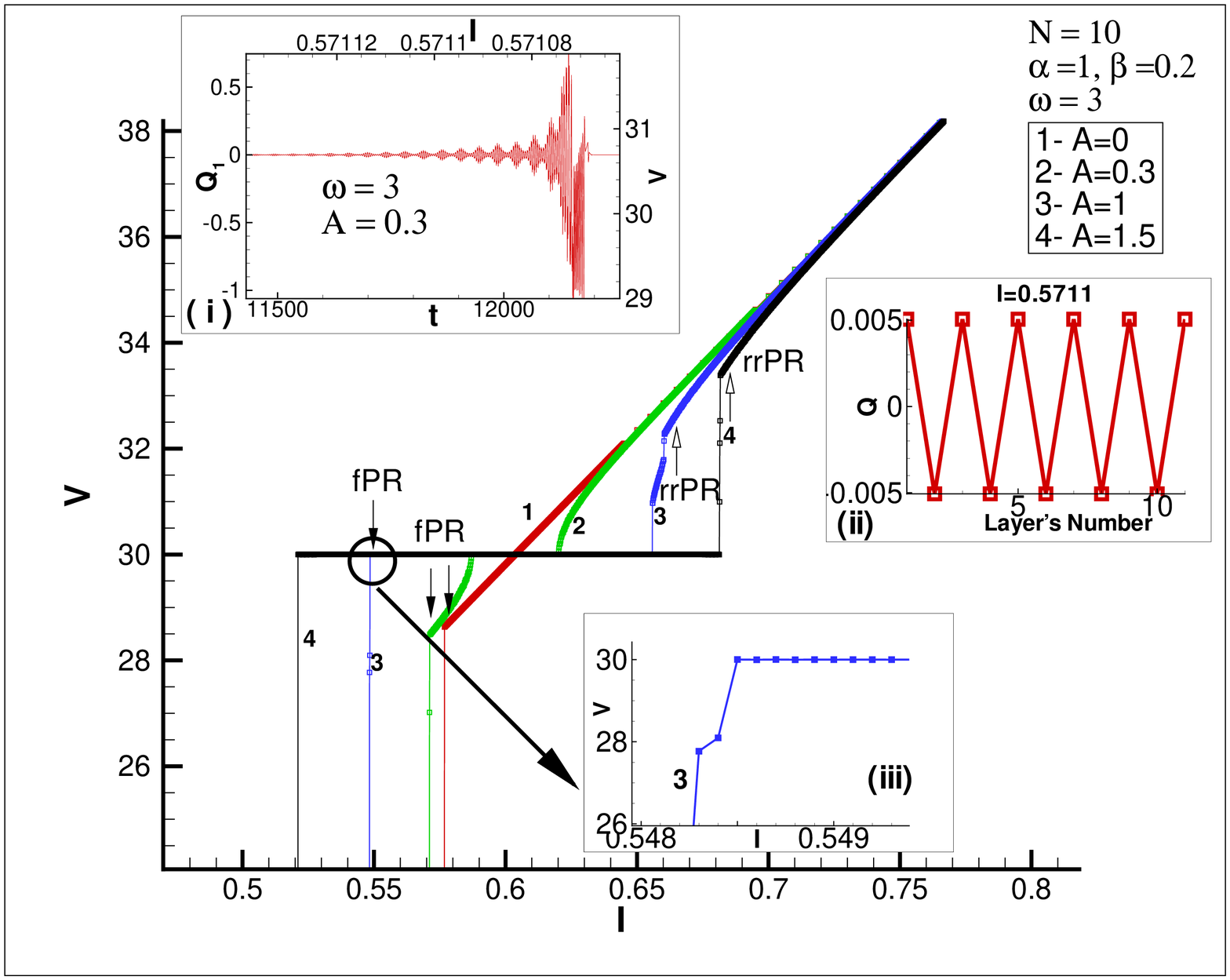}
\caption{CVC of the stack with 10 coupled JJ without irradiation (curve 1) and under radiation with frequency $\omega=3$ and
amplitude $A=0.3$ (curve 2);  $A=1$ (curve 3); $A=1.5$ (curve 4). Inset (i) illustrate the modulation of the charge oscillation and inset (ii) shows the charge distribution among the layers at $A=0.3$. Inset (iii) enlarges the fPR region in the curve 3.}
 \label{2}
\end{figure}
Let us see the influence of the external radiation on the PR by the increase of amplitude of the radiation at fixed frequency. In this present work, we describe the case $\omega > 2\omega_{LPW}$, where the SS is above PR region in CVC. Fig.~\ref{2} shows 4 CVCs of a stack with 10 coupled JJs: without
irradiation (curve 1) and under radiation with $\omega=3$,
$A = 0.3$ (curve 2), $A = 1$ (curve 3) and $A = 1.5$ (curve 4). At $\omega=0$, PR is characterized by breakpoint current $I_{bp}\simeq 0.5772$, and breakpoint voltage $V_{bp} \simeq 28.65$, corresponding to the Josephson frequency $\omega_J=2.865$.

As shown in the figure, the first SS is developed
on the outermost branch of CVC in the hysteresis region
at $V = \omega_J*N = 30$. It was shown in Ref. \cite{Gaafar2012} that an increase of the amplitude of irradiation leads to the appearance of an additional PR before SS called radiation related parametric resonance (rrPR). Filled arrows indicate the positions corresponding to the appearance of a fundamental PR (fPR) in the stack, which
realizes without radiation. Hollow arrow indicates the  rrPR caused by irradiation. Inset (i) illustrate the modulation of the charge oscillations at $A=0.3$, while inset (ii) shows the charge distribution among the S-layers in the growing region presented in inset (i). The charges on the neighbor layers are equal in magnitude and opposite in sign. This is true for all adjacent layers and corresponds to the $\pi$-mode. So, the $\pi$-mode is survived under radiation with A = 0.3. Results of FFT analysis of the time
dependence of voltage V(t) in each JJ and charge Q(t) in
each S-layer show that this modulation is due to beating
between the external and Josephson frequencies. We see in the inset (iii) an  enlarged part of CVC corresponding to the appearance of fPR region in curve 3.

The irradiation changes the shape of CVC as it is seen in the curve 3 at $\omega=3$ and $A=1$. At these parameters, as was mentioned above, the rrPR appears
in the stack before SS (shown by hollow arrow). For this case, the charge-time dependencies in the fPR and rrPR regions are shown in Fig. 3.
\begin{figure}[!ht]
 \centering
\includegraphics[height=50mm]{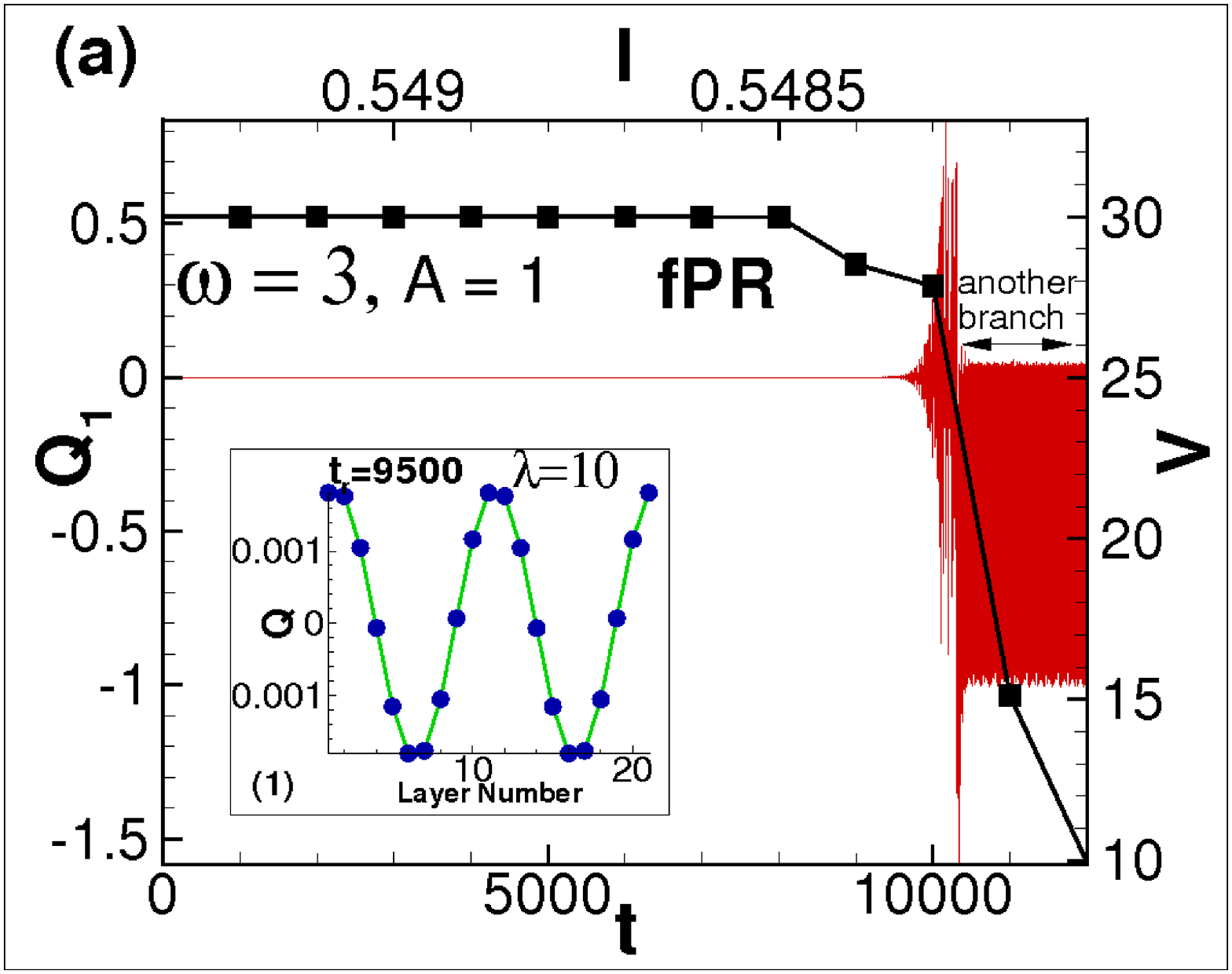}
\includegraphics[height=50mm]{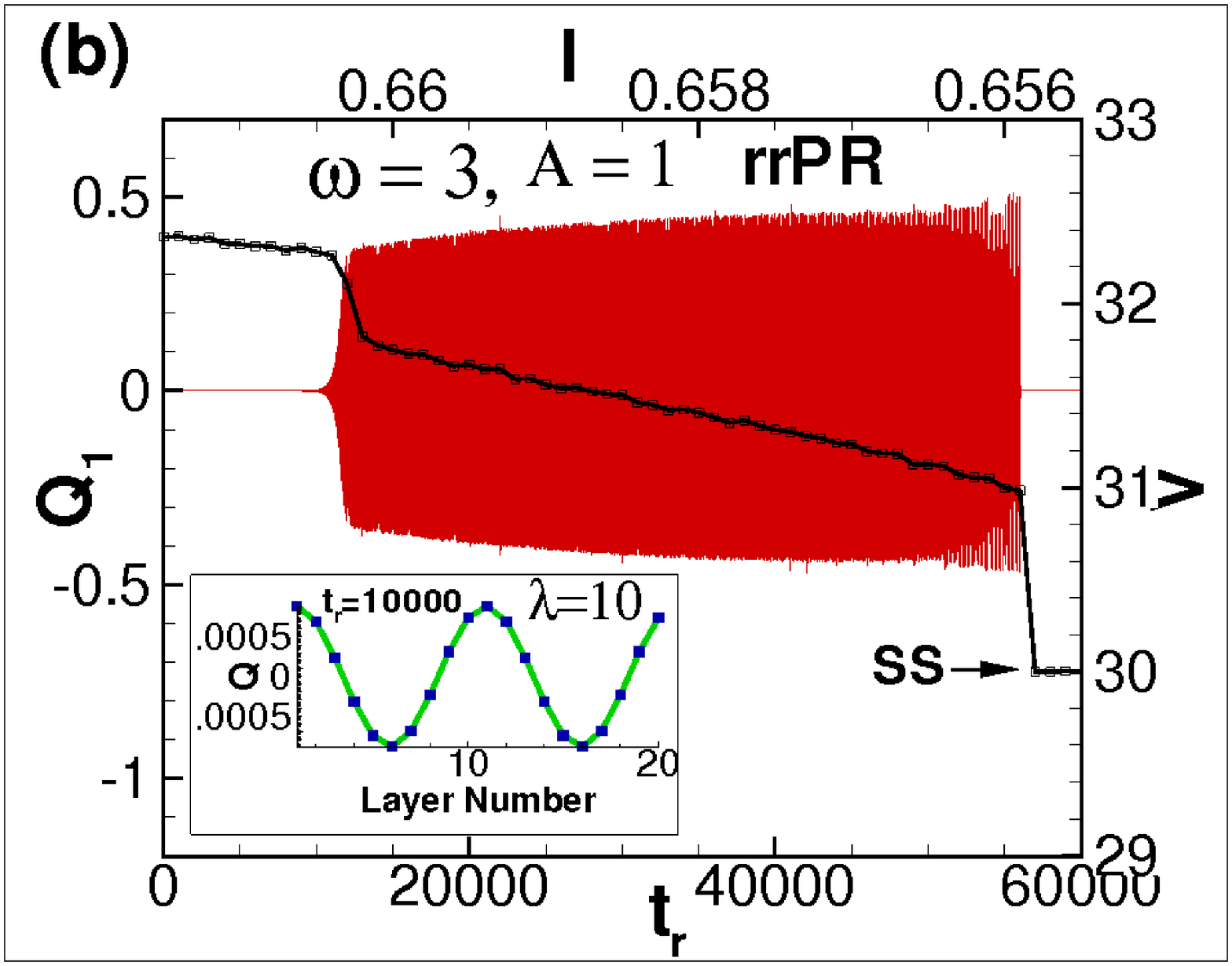}
\caption{Charge time-dependence at $\omega=3$ and $A=1$ for (a) fPR region; (b) rrPR region. The insets show the charge distributions among the S-layers in the growing region of resonances.}
 \label{3}
\end{figure}
Fig. 3(a) demonstrates the time dependence of the charge in the fPR region for the first S-layer at $\omega=3$ and $A=1$ which is combined with CVC of the outermost branch. The filled squares
mark the bias current steps in CVC. The inset shows the charge distribution among the layers, illustrates that the wavelength of the LPW is changed with increasing the amplitude to $\lambda=10 d$. In Fig. 3(b) the time dependence of the charge in the rrPR region is introduced. We see clearly the correlation between the charge-time dependence and CVC. The inset shows that the wavelength of LPW at this rrPR is also $\lambda=10 d$.
\section{Conclusions}
We demonstrated different effects of external electromagnetic radiation on the parametric resonance of coupled Josephson junctions. Approaching of the fundamental parametric resonance by the decreasing  of the  radiation's frequency  shrinks the width of this resonance. An increase in the amplitude of radiation leads to (a) changing the wavelength of longitudinal plasma wave along the stack, (b) appearance of additional parametric resonance above Shapiro step.
\section{Acknowledgements}
The support by scientific agreement between Egypt and JINR, Dubna, Russia is acknowledged.

\end{document}